\documentclass[reprint,amsmath,amssymb,aps,footinbib]{revtex4-1}  
\usepackage{dcolumn}
\usepackage{color}
\usepackage{bm}
\usepackage{xspace}
\usepackage{graphicx}           

\newcommand{\be}{\begin{equation}}
\newcommand{\ee}{\end{equation}}
\newcommand{\bea}{\begin{eqnarray}}
\newcommand{\eea}{\end{eqnarray}}
\newcommand{\beaa}{\begin{eqnarray*}}
\newcommand{\eeaa}{\end{eqnarray*}}
\newcommand{\alphaI}{{$\alpha$-(BEDT-TTF)$_2$I$_3$}\xspace}

\begin{document}

\title{
  Time-Reversal Symmetry-Breaking Flux State in an Organic Dirac Fermion System
}

\author{Takao Morinari}
 \email{morinari.takao.5s@kyoto-u.ac.jp}
 \affiliation{Course of Studies on Materials Science,
  Graduate School of Human and Environmental Studies,
  Kyoto University, Kyoto 606-8501, Japan
}

\date{\today}

\begin{abstract}
  We investigate symmetry breaking in the Dirac fermion phase
  of the organic compound \alphaI under pressure,
  where BEDT-TTF denotes bis(ethylenedithio)tetrathiafulvalene.
  The exchange interaction resulting from inter-molecule Coulomb repulsion
  leads to broken time-reversal symmetry and particle-hole symmetry,
  while preserving translational symmetry.
  The system breaks time-reversal symmetry by creating fluxes in the unit cell.
  This symmetry-broken state exhibits a large Nernst signal as well as thermopower.
  We compute the Nernst signal and thermopower,
  demonstrating their consistency with experimental results.
\end{abstract}

\maketitle

\section{Introduction}
Organic charge-transfer salt, \alphaI, has been extensively studied
as a quasi-two-dimensional Dirac fermion system \cite{Katayama2006,Kobayashi2007,Kajita2014}.
Here, BEDT-TTF refers to\\
bis(ethylenedithio)tetrathiafulvalene.
The extended Hubbard model, incorporating the pressure dependence
in the transfer energies obtained
from X-ray diffraction experiments \cite{Mori1984,Mori1999,Kondo2005},
predicts a two-dimensional Dirac fermion spectrum with charge disproportionation
under high pressures \cite{Kobayashi2007}, a result corroborated
by first principles calculations \cite{Ishibashi2006,Kino2006}.
The presence of the Dirac fermion spectrum is evident through
the large negative interlayer magnetoresistance, attributed
to the zero-energy Landau level of the Dirac fermions \cite{Osada2008,Tajima2009}.
Additionally, the phase of the Dirac fermions is confirmed
through Shubnikov-de Haas oscillations in hole-doped samples placed
on polyethylene naphthalate substrates \cite{Tajima2013}.

Recently, research on \alphaI has reached a turning point with theoretical results
suggesting broken time-reversal and inversion symmetry \cite{Morinari2020},
and experimental findings indicating that the system exhibits
characteristics of a three-dimensional Dirac semimetal
at low temperatures \cite{Tajima2023b}.
The observation of a peak in the interlayer magnetoresistance at low temperatures
indicates phase-coherent interlayer tunneling,
implying a three-dimensional electronic structure \cite{Tajima2023}.
Furthermore, the detection of negative magnetoresistance
and the planar Hall effect provide supporting evidence for \alphaI
being a Dirac semimetal \cite{Tajima2023b,Uji2024}.

In this paper, we investigate the origin of broken time-reversal symmetry
and inversion symmetry in \alphaI.
We find that the symmetry breaking arises from a flux state.
The unit cell of \alphaI contains four BEDT-TTF molecules,
as depicted in Fig.~\ref{fig:t_def}, denoted by A, A$^{\prime}$, B, and C,
and inversion symmetry exists between A and A$^{\prime}$
molecules\cite{Mori1984,Kakiuchi2007,Piechon2013}.
We demonstrate that this inversion symmetry is broken
and the exchange interaction induces asymmetry
in the phases of the hopping matrix elements between A and A$^{\prime}$ molecules,
consequently breaking time-reversal symmetry and particle-hole symmetry.

While direct verification of broken time-reversal symmetry
is challenging due to the system being in a pressure cell,
we show that the asymmetry of the Dirac point locations concerning the Fermi energy
results in a substantial Nernst signal with non-vanishing thermopower.
This result is in good agreement with experimental observations \cite{Konoike2013}.

The rest of the paper is organized as follows:
In Sec.~\ref{sec:Model}, we present the Hamiltonian
for \alphaI and introduce bond mean fields and charge mean fields.
In Sec.~\ref{sec:flux}, we demonstrate how the mean field state breaks
time-reversal symmetry by creating fluxes within a unit cell
while maintaining translational symmetry.
To investigate how the broken time-reversal symmetry can be experimentally confirmed,
we compute the thermopower and the Nernst signal in Sec.~\ref{sec:transport}.
We demonstrate that the system displays a significant Nernst signal,
several times greater than the thermopower,
consistent with experimental observations\cite{Konoike2013}.

\section{Model and Order Parameters}
\label{sec:Model}
In our theoretical study of \alphaI, we primarily examine
the conduction layer composed of BEDT-TTF molecules.
Recent experiments investigating the azimuthal angular dependence
of interlayer magnetoresistance\cite{Tajima2023} have provided key insights.
Specifically, the interlayer tunneling energy is estimated to be approximately 1 meV,
inferred from the peak width observed when the magnetic field is aligned within the plane.

Given that the transfer integrals between BEDT-TTF molecules within the plane
are two orders of magnitude greater\cite{Mori1984,Kondo2005}
than this interlayer tunneling energy,
it becomes reasonable to adopt a two-dimensional model for our analysis.
This approach effectively captures the dominant in-plane interactions
while acknowledging the quasi-two-dimensional nature of the actual system.

\begin{figure}[htbp]
  \includegraphics[width=0.7 \linewidth]{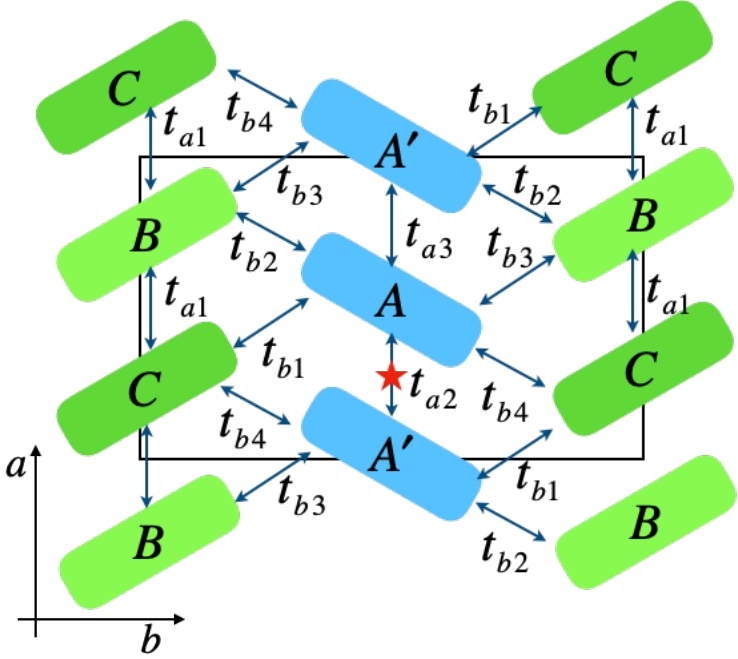}
  \caption{
    \label{fig:t_def}
    (Color online)
    Conducting plane of BEDT-TTF molecules in \alphaI and the transfer energies
    between the molecules. The solid rectangle represents the unit cell.
    The crystal axes $a$ and $b$ are also shown.
    The red star indicates the inversion center between molecule A and molecule A$^\prime$.
    }
\end{figure}

The transfer energies between molecules in the conduction plane of \alphaI
are depicted in Fig.~\ref{fig:t_def}.
The Hamiltonian describing electron hopping is represented by
the following Hamiltonian:
\be
   {H_0} = \sum\limits_{\left\langle {i,j} \right\rangle } {\sum\limits_{\alpha ,\beta }
     {\sum\limits_\sigma  {{t_{i\alpha ,j\beta }}c_{i\alpha \sigma }^\dag {c_{j\beta \sigma }}} } }.
\ee
Here, $\langle {i,j} \rangle$
represents the nearest neighbor unit cells,
and the indices $\alpha$ and $\beta$ correspond
to molecules A, A$^{\prime}$, B, and C.
Hereafter, we will refer to A, A$^{\prime}$, B, and C as 1, 2, 3, and 4, respectively.
The operator
${c_{i\alpha \sigma }^\dag }$ ( ${{c_{i \alpha \sigma }}}$ )
creates (annihilates) an electron with spin $\sigma$
at molecule $\alpha$
in $i$-th unit cell.

\begin{figure}[htbp]
  \includegraphics[width=1 \linewidth]{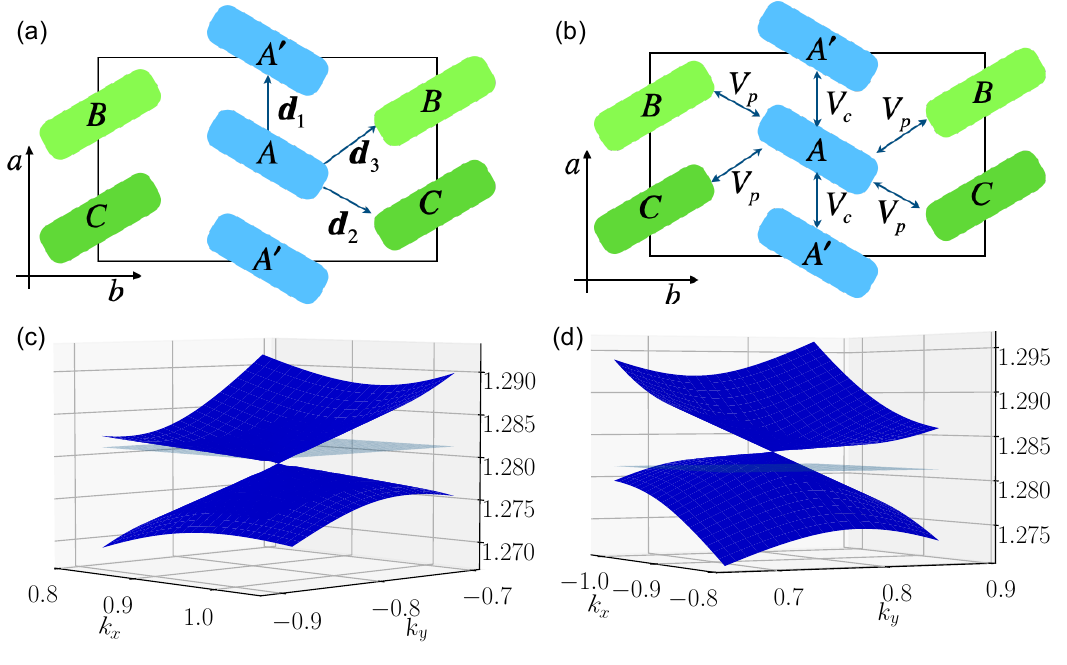}
  \caption{
    \label{fig:d_V_disp}
    (Color online)
    (a) Definition of three vectors, ${\bm{d}}_1$, ${\bm{d}}_2$, and ${\bm{d}}_3$
    within the unit cell.
    (b) The nearest-neighbor interactions
    between molecule A and the other molecules.
    $V_c$ represents the interaction between A and A$^{\prime}$,
    while $V_p$ corresponds to the interactions between A and B, and A and C.
    The same set of interactions is also present
    for molecules A$^{\prime}$, B, and C.
    (c) and (d): The energy dispersions around the Dirac points
    obtained by the mean field calculation with $x=1$ and $P=0.8$~GPa.
    The energy dispersions are plotted around the Dirac points
    at (c) ${\bm k}_D^{(1)}=(0.939,-0.806)$ 
    and (d) ${\bm k}_D^{(2)}=(-0.908,0.760)$.
    The thin horizontal plane represents the Fermi energy.
    In (c), the Dirac point is located below the Fermi energy,
    while in (d), the Dirac point is located above the Fermi energy.
}
\end{figure}

Including the interaction terms to $H_0$, 
the conduction layer in \alphaI is described by the following
extended Hubbard model\cite{Kobayashi2007}:
\bea
H &=& \sum\limits_{{\bm{k}},\alpha ,\beta ,\sigma }
{{\varepsilon _{\alpha \beta }}\left( {\bm{k}} \right)c_{{\bm{k}}\alpha \sigma }^\dag
  {c_{{\bm{k}}\beta \sigma }}}  + U\sum\limits_{j,\alpha } {c_{j\alpha  \uparrow }^\dag
  c_{j\alpha  \downarrow }^\dag {c_{j\alpha  \downarrow }}{c_{j\alpha  \uparrow }}}
\nonumber \\
& & + \sum\limits_{i,j,\alpha ,\beta ,\sigma ,\sigma '} {{V_{i\alpha ,j\beta }}
  c_{i\alpha \sigma }^\dag c_{j\beta \sigma '}^\dag {c_{j\beta \sigma '}}{c_{i\alpha \sigma }}}.
\label{eq:H}
\eea
The operator ${c_{{\bm{k}}\alpha \sigma }^\dag }$ ( ${{c_{{\bm{k}}\alpha \sigma }}}$ )
creates (annihilates) an electron with the Bloch state ${\bm{k}}$ and spin $\sigma$
at molecule $\alpha$.
We assume that the system is subjected to uniaxial pressure
along the $a$-axis \cite{Kondo2005}.
${\varepsilon _{\alpha \beta }}\left( {\bf{k}} \right)$ are 
given by \cite{Kobayashi2007}
\bea
    {\varepsilon _{12}}\left( {\bm{k}} \right)
    &=& {t_{a3}}{e^{ - i{\bm{k}} \cdot {{\bm{d}}_1}}} + {t_{a2}}{e^{i{\bm{k}}
        \cdot {{\bm{d}}_1}}},
    \label{eq:transfer_energy1}
    \\
    {\varepsilon _{13}}\left( {\bm{k}} \right) &=& {t_{b3}}{e^{ - i{\bm{k}}
        \cdot {{\bm{d}}_3}}} + {t_{b2}}{e^{i{\bm{k}} \cdot {{\bm{d}}_2}}},\\
    {\varepsilon _{14}}\left( {\bm{k}} \right) &=& {t_{b4}}{e^{ - i{\bm{k}}
        \cdot {{\bm{d}}_2}}} + {t_{b1}}{e^{i{\bm{k}} \cdot {{\bm{d}}_3}}},\\
    {\varepsilon _{23}}\left( {\bm{k}} \right) &=& {t_{b2}}{e^{ - i{\bm{k}}
        \cdot {{\bm{d}}_2}}} + {t_{b3}}{e^{i{\bm{k}} \cdot {{\bm{d}}_3}}},\\
    {\varepsilon _{24}}\left( {\bm{k}} \right) &=& {t_{b1}}{e^{ - i{\bm{k}}
        \cdot {{\bm{d}}_3}}} + {t_{b4}}{e^{i{\bm{k}} \cdot {{\bm{d}}_2}}},\\
    {\varepsilon _{34}}\left( {\bm{k}} \right) &=& {t_{a1}}{e^{i{\bm{k}}
        \cdot {{\bm{d}}_1}}} + {t_{a1}}{e^{ - i{\bm{k}} \cdot {{\bm{d}}_1}}}.
    \label{eq:transfer_energy6}    
    \eea
    Here, the transfer energies, $t_\lambda$ ($\lambda = a1, a2, ...$),
    are pressure-dependent and can be expressed as follows:
    \[
      {t_\lambda } = {C_\lambda }\left( {1 + {b_\lambda }P} \right).
      \]
      The numerical coefficients $C_\lambda$ and $b_\lambda$ are as follows:
      For the molecule stacking direction:
      $C_{a1} = -0.028$, $b_{a1} = 0.89$,
      $C_{a2} = 0.048$, $b_{a2} = 1.67$,
      $C_{a3} = -0.020$, $b_{a3} = -0.25$.
      For the other directions:
      $C_{b1} = 0.123$, $b_{b1} = 0$,
      $C_{b2} = 0.140$, $b_{b2} = 0.11$,
      $C_{b3} = -0.062$, $b_{b3} = 0.32$,
      $C_{b4} = -0.025$, $b_{b4} = 0$ \cite{Kobayashi2007}.
      $C_\lambda$ is given in units of eV, and the pressure $P$ is in units of GPa.
        Hereafter, we take eV as the unit of energy.
      These values are derived using an extrapolation formula \cite{Kobayashi2004},
      which is based on band calculations at ambient pressure \cite{Mori1999}
      and crystal structure analysis under pressure\cite{Kondo2005}.
      As shown in Fig.~\ref{fig:d_V_disp}(a),
      the vectors ${\bm{d}}_1$, ${\bm{d}}_2$, and ${\bm{d}}_3$ are defined as follows:
    \be
       {{\bm{d}}_1} = \left( {0,\frac{a}{2}} \right),
       {{\bm{d}}_2} = \left( {\frac{b}{2}, - \frac{a}{4}} \right),
       {{\bm{d}}_3} = \left( {\frac{b}{2},\frac{a}{4}} \right),
       \ee
       where $a$ and $b$ are the lattice constants
       along the $a$-axis and $b$-axis, respectively.
We note that $a=9.19 \AA$, $b=10.80 \AA$, and $c=17.39 \AA$ at ambient pressure.

The second term on the right-hand side of Eq.~(\ref{eq:H})
represents the on-site Coulomb interaction,
while the third term describes the nearest-neighbor interactions
between different molecules.
The parameters, $V_{i\alpha ,j\beta }$, can take values of either $V_c$ or $V_p$,
as shown in Fig.~\ref{fig:d_V_disp}(b),
depicting the interactions between molecule A and other molecules.
The same set of interactions is also present among the other molecules.

Now, we apply a mean-field approximation to Eq.~(\ref{eq:H}).
The charge order at molecule $\alpha$ with spin $\sigma = \uparrow, \downarrow$
is defined as follows:
\be
{{n_{\alpha \sigma }}}
= \frac{1}{N}\sum\limits_{\bm{k}} {\left\langle {c_{{\bm{k}}\alpha \sigma }^\dagger
{c_{{\bm{k}}\alpha \sigma }}} \right\rangle },
\ee
Here, $N$ is the number of unit cells.
This charge order plays a pivotal role in describing the insulating state
at ambient pressure \cite{Seo2000,Kino1996}.
  A thorough mean field calculation was conducted by Seo \cite{Seo2000},
  examining the stability of various symmetry-broken phases.
  Utilizing parameters $U=0.4$, $V_c=0.17$, and $V_p=0.05$,
  the mean field calculation successfully reproduces Seo's findings
  regarding stripe charge order at ambient pressure \cite{Kobayashi2007}.
  Under conditions of high pressure, the relative strength of short-range Coulomb interactions
  to transfer energies diminishes, leading to the replacement of the stripe charge order
  by a Dirac fermion state with charge disproportionation
  above 0.43 GPa \cite{Kobayashi2011a}.
  At a pressure of 1.5 GPa, the experimental condition investigated
  by Konoike et al. \cite{Konoike2013},
  the emergence of charge density wave and spin density wave phases can be confidently excluded.

The broken time-reversal symmetry and inversion symmetry
are described by the following bond order parameter\cite{Morinari2020}:
\be
{\chi _{\alpha \sigma ,\beta \sigma ', \pm }} 
= \frac{1}{N}\sum\limits_{\bm{k}} {{e^{ - i{\bm{k}} 
\cdot {\bm{d}}_{\alpha \beta }^{\left(  \pm  \right)}}}
\left\langle {c_{{\bm{k}}\alpha \sigma }^\dagger 
{c_{{\bm{k}}\beta \sigma '}}} \right\rangle }.
\ee
We need to distinguish two bonds connecting molecule $\alpha$ and $\beta$.
The vectors connecting molecule $\alpha$ and $\beta$ are denoted
by ${\bm{d}}_{\alpha \beta }^{\left(  \pm  \right)}$.
For example, ${\bm{d}}_{13}^{\left(  +  \right)} = {{\bm{d}}_2}$
and ${\bm{d}}_{13}^{\left(  -  \right)} = - {{\bm{d}}_3}$.
Regarding the interaction parameters,
we adopt the assumption\cite{Kobayashi2007} that
$U=0.4x$, $V_c=0.17x$, and $V_p=0.05x$,
with these values expressed in units of eV.
Here, the parameter $x$ controls the strength of the interaction.
The case where $x=1$ results in charge disproportionation,
which is consistent with the findings of NMR experiments\cite{Moroto2004}.
We conduct self-consistent calculations
for ${\chi _{\alpha \sigma ,\beta \sigma ', \pm }}$
and $n_{\alpha \sigma}$.

\section{Broken Time-Reversal Symmetry and Inversion Symmetry}
\label{sec:flux}
The mean-field state breaks both time-reversal symmetry and inversion symmetry.
Broken inversion symmetry is evident from the energy dispersion,
as shown in Fig.~\ref{fig:d_V_disp}(c) and (d),
where we take $P=0.8$~GPa and $x=1$.
In the $k_x$-$k_y$ plane, there exist two Dirac points.
One Dirac point is located at ${\bm k}_D^{(1)}=(0.939,-0.806)$,
and the other is at ${\bm k}_D^{(2)}=(-0.908,0.760)$.
From these values, it is clear that inversion symmetry is broken since,
if inversion symmetry were present, ${\bm k}_D^{(1)}=-{\bm k}_D^{(2)}$.

The energies at the Dirac points are as follows:
${E_{{\bm k}_D^{(1)}}} = {\rm{1.279}}$ and
${E_{{\bm k}_D^{(2)}}} = {\rm{1.283}}$,
while the Fermi energy is ${E_F} = {\rm{1.281}}$.
There are electrons around the Dirac point at ${\bm k}_D^{(1)}$,
and there are holes around the other Dirac point at ${\bm k}_D^{(2)}$.
Note that  the particle-hole symmetry is broken
because
$E_F - {E_{{\bm k}_D^{(1)}}} \neq {E_{{\bm k}_D^{(2)}}}-E_F$.

\begin{figure}[htbp]
  \includegraphics[width=1.0 \linewidth, angle=0]{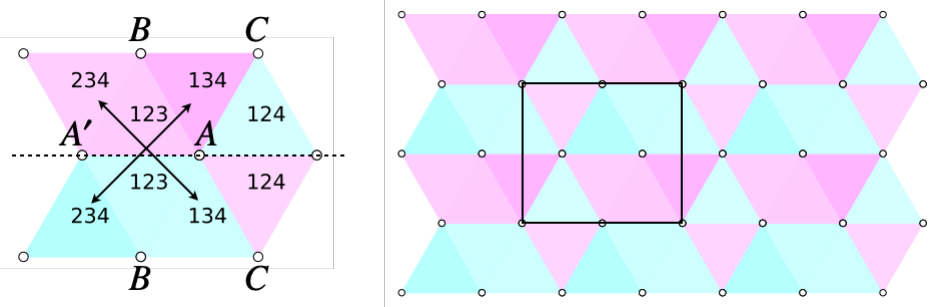}
  \caption{
    \label{fig:flux_state}
    (Color online)    
    Left panel: The flux configuration in the unit cell,
    with open circles representing molecules.
    Each plaquette is labeled with three numbers denoting $\alpha$, $\beta$,
    and $\gamma$ of ${\chi _{\alpha \beta \gamma }}$.
    Plaquettes colored in pink denote positive flux values,
    while those in blue signify negative flux values.
    Subtle variations in hue represent deviations from $\pm \pi$.
    The pairs of plaquettes connected by arrows indicate
    that the fluxes of these two plaquettes have opposite signs
    and thus cancel each other out.
    Right panel: Flux patterns in unit cells are shown, exhibiting translational symmetry.
    The solid rectangle represents the unit cell.
  }
\end{figure}

The broken time-reversal symmetry state is the flux state,
as described below.
We find that the spin degeneracy is not lifted,
and the $\sigma=\uparrow$ states and
$\sigma=\downarrow$ states remain decoupled:
${\chi _{\alpha \uparrow ,\beta \uparrow ', \pm }}
={\chi _{\alpha \downarrow ,\beta \downarrow ', \pm }}
$,
${\chi _{\alpha \uparrow ,\beta \downarrow ', \pm }}=0$,
and
${\chi _{\alpha \downarrow ,\beta \uparrow ', \pm }}=0$.
As a result, the following Hamiltonian is added to the kinetic energy term:
\bea
    {H_\chi } &=&
    - \frac{1}{2}\sum\limits_{{\bf{k}},\alpha \neq \beta ,\sigma }
    \sum\limits_{s =  \pm } {{V_{\alpha \beta }}}
      \left( {\chi _{\alpha  \uparrow ,\beta  \uparrow ,s}^*
        + \chi _{\alpha  \downarrow ,\beta  \downarrow ,s}^*} \right)
      \nonumber \\
      & & \times
                {e^{i{\bf{k}} \cdot {\bf{d}}_{\alpha \beta }^{\left(  s  \right)}}}
                c_{{\bf{k}}\alpha \sigma }^\dag {c_{{\bf{k}}\beta \sigma }}.
   \eea
   Here, $V_{\alpha \beta}$
   represents the interaction parameter between molecule
   $\alpha$ and molecule $\beta$.
   In terms of ${\bf{d}}_{\alpha \beta }^{\left( s \right)}$,
   the term ${\varepsilon _{\alpha \beta }}\left( {\bf{k}} \right)$ is
   rewritten as:
   \be
     {\varepsilon _{\alpha \beta }}\left( {\bf{k}} \right) = \sum\limits_{s =  \pm } {{t_{\alpha \beta s}}{e^{i{\bf{k}} \cdot {\bf{d}}_{\alpha \beta }^{\left( s \right)}}}},
     \ee
     with ${t_{13 + }} = {t_{b2}}$, ${t_{13 - }} = {t_{b3}}$, and so forth.
     Upon including $H_\chi$,
     the term ${{t_{\alpha \beta s}}}$ is replaced by
     $t_{\alpha \beta s}^{{\rm{eff}}}$, where
   \be
   t_{\alpha \beta s}^{{\rm{eff}}} = {t_{\alpha \beta s}} - \frac{1}{2}{V_{\alpha \beta }}\left( {\chi _{\alpha  \uparrow ,\beta  \uparrow ,s}^* + \chi _{\alpha  \downarrow ,\beta  \downarrow ,s}^*} \right).
   \ee
An effective hopping parameter, $t_{\alpha \beta  \pm }^{{\rm{eff}}}$,
is introduced, which combines the original hopping parameter $t_{\lambda}$
with the bond order parameter ${\chi _{\alpha \sigma ,\beta \sigma ', \pm }}$.
Using $t_{\alpha \beta  \pm }^{{\rm{eff}}}$,
we define the product of three effective hopping parameters
around each triangular plaquette.
For instance,
\be
  {\chi _{124}^{(-)}} = t_{14 + }^{{\rm{eff}}}t_{42 - }^{{\rm{eff}}}t_{21 - }^{{\rm{eff}}}.
  \ee
  We define this product in the counterclockwise direction.
  Additionally, there is another plaquette involving
  molecules 1, 2, and 4.
  That is,
  \be
     {\chi _{124}^{(+)}} = t_{12 + }^{{\rm{eff}}}t_{24 - }^{{\rm{eff}}}t_{41 + }^{{\rm{eff}}}.
     \ee
       In Fig.~\ref{fig:flux_state}, left panel,
     a horizontal dashed line divides the unit cell into an upper part and a lower part. 
     The $\pm$ sign in $\chi _{\alpha \beta \gamma}^{(\pm)}$
     refers to the two plaquettes within the unit cell.
     Specifically, the $+$ sign corresponds to the plaquettes located in the upper region,
     while the $-$ sign corresponds to those in the lower region.
     The argument of the complex number
     ${\chi _{\alpha \beta \gamma }^{(\pm)}}$
     is denoted by ${\Phi _{\alpha \beta \gamma }^{(\pm)}}$
     and is defined by
     \be
        {\chi _{\alpha \beta \gamma }^{(\pm)}}
        = \left| {{\chi _{\alpha \beta \gamma }^{(\pm)}}}
        \right|\exp \left( {i{\Phi _{\alpha \beta \gamma }^{(\pm)}}} \right).
        \ee
${\Phi _{\alpha \beta \gamma }^{(\pm)}}$ takes either 0 or $\pi$
for the non-interacting case \cite{Piechon2015}.

  Time-reversal symmetry is broken by generating fluxes 
  at each triangle within the unit cell, with values neither 0 nor $\pm \pi$, 
  as depicted in Fig.~\ref{fig:flux_state} (left panel). 
  Importantly, both spin-up and spin-down states contribute to the same flux pattern,
  and the state does not involve any magnetization.
The flux pattern of ${\Phi _{\alpha \beta \gamma }^{(\pm)}}$ is
depicted in Fig.~\ref{fig:flux_state},
where plaquettes with pink color indicate ${\Phi _{\alpha \beta \gamma }^{(\pm)}}>0$,
and blue color indicates ${\Phi _{\alpha \beta \gamma }^{(\pm)}}<0$.
Subtle variations in hue represent deviations from $\pm \pi$.
Within each unit cell, the fluxes collectively sum to zero: 
${\chi _{124}^{(+)}}$ offsets ${\chi _{124}^{(-)}}$. 
Similarly, ${\chi _{123}^{(+)}}$ offsets ${\chi _{123}^{(-)}}$. 
Concomitantly, ${\chi _{134}^{(\pm)}}$ offsets ${\chi _{234}^{(\mp)}}$.  

\begin{figure}[htbp]
  \includegraphics[width=0.90 \linewidth, angle=0]{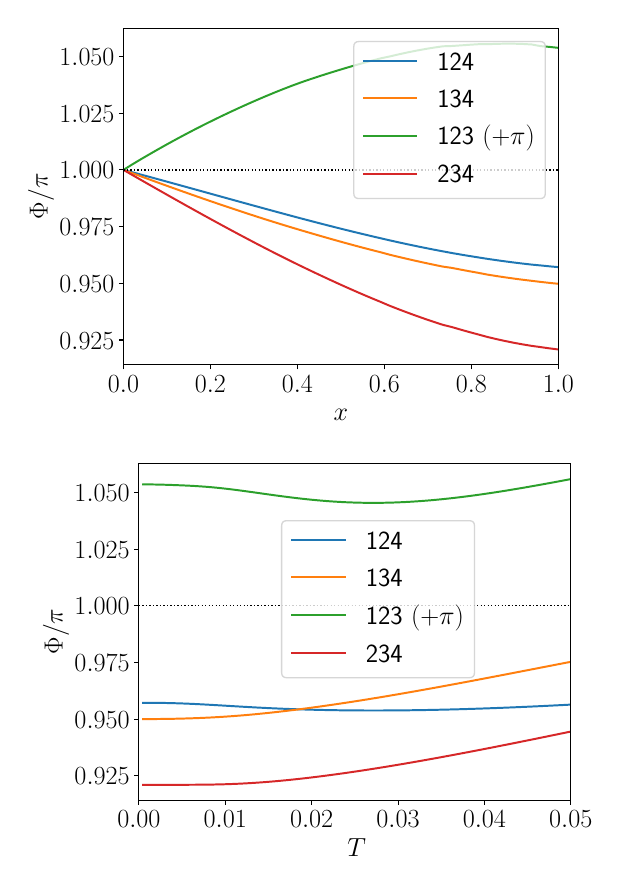}
  \caption{
    \label{fig:flux_state_values}
    (Color online)
    Upper panel:
    Dependence of the flux values on the interaction.
    Only flux values where ${\Phi _{\alpha \beta \gamma }}>0$ are shown.
    The value of ${\Phi _{123}}$ is shifted by $\pi$.
    Lower panel:
    Temperature dependence of the flux values.
  }
\end{figure}

While time-reversal symmetry is broken within the unit cell,
the translational symmetry remains unbroken,
as shown in the right panel of Fig.~\ref{fig:flux_state}.
We note that a similar time-reversal symmetry breaking is discussed
on the high-$T_c$ cuprates\cite{Varma1997}.
The dependence of the flux values on the interaction strength
is demonstrated in
the upper panel of Fig.~\ref{fig:flux_state_values}.  
We confirmed from the calculation with $U=0$ (not shown)
that the nearest-neighbor interactions $V_c$ and $V_p$ play
a central role in breaking time-reversal symmetry.
Regarding the temperature dependence of the flux values,
the lower panel of Fig.~\ref{fig:flux_state_values} 
presents the results.
Notably, there is no phase transition within the temperature range
displayed in this figure.
Time-reversal symmetry is already broken at high temperatures.

From a detailed analysis of $t_{\alpha \beta  \pm }^{{\rm{eff}}}$,
we find that the breaking of time-reversal symmetry
originates from the symmetry breaking in the hopping between A and A$^{\prime}$.
That is,
\be
t_{12+}^{{\rm{eff}}} \neq \left( t_{21-}^{{\rm{eff}}} \right)^*.
\ee
We note that the symmetry breaking in $n_A=n_{A\uparrow}+n_{A\downarrow}$
and $n_{A^{\prime}}=n_{{A^{\prime}}\uparrow}+n_{{A^{\prime}}\downarrow}$
leads to stripe charge order under ambient pressure \cite{Seo2000,Kino1996,Takahashi2003}.
However, at high pressure, this symmetry breaking is replaced by
the breaking of time-reversal and inversion symmetries.

  We now address the reliability of our mean field calculations. 
  Given that the system under study is strongly correlated, 
  concerns about the impact of electronic correlations are valid. 
  These effects were thoroughly investigated 
  in our previous work\cite{Unozawa2020}
  using slave-rotor theory\cite{Florens2004}. 
  We demonstrated that strong coupling effects significantly 
  reduce the Fermi velocity, which is a critical property of Dirac fermions,
  aligning our findings with renormalization group theory\cite{Tang2018}.
  Notably, the system transitions to an insulating state
  at a finite pressure of approximately 0.6 GPa.
  Furthermore, we compared the pressure-dependent Fermi velocity with experimental data
  derived from Shubnikov-de Haas oscillation analysis in doped samples.
  The results exhibited good agreement with these experimental observations.
  However, it is important to recognize that the influence of strong coupling effects
  diminishes at higher pressures.
  This reduction is due to the decreased ratio of on-site Coulomb repulsion
  to transfer energies as pressure increases.
  At pressures of 0.8 GPa or higher, the effect of
  the strong electronic correlation becomes negligible.

As for the stability of the flux state,
we comment on the pivotal role of off-site Coulomb $V$ interactions
in the emergence of the flux state in \alphaI.
The stability of the flux state, as compared to the charge-ordered state, 
can be attributed to the specific characteristics of the density of states (DOS)
at the Fermi energy. 
Notably, the DOS vanishes at the Dirac point, where the Fermi energy lies
in the absence of the interactions. 
This phenomenon significantly impedes the stabilization of a charge-ordered state. 
In contrast, the flux state, which originates from phase modifications
in the hopping parameters, exhibits a relatively independent stability of the DOS. 
This distinction is crucial for understanding the preferential formation
of the flux state under these conditions.

Now we discuss the distinctions between our flux state 
and the topological Mott insulators\cite{Raghu2008}. 
In the model of topological Mott insulators,
a flux value of $\pm \pi/2$ is specifically chosen to create a gap 
within the Dirac fermion spectrum. 
By contrast, our study investigates a flux state characterized 
by a non-zero flux value, which does not inherently result in a gap. 
This difference is primarily due to the vanishing DOS in the non-interacting system, 
a factor that significantly influences the behavior 
and properties of the flux state under consideration.

\section{Thermopower and Nernst Signal}
\label{sec:transport}
The experimental verification of broken time-reversal symmetry is a crucial question
to address.
However, since this state is realized under high pressure,
it requires the use of a pressure cell, making direct verification extremely difficult.
Nevertheless, the presence of broken time-reversal symmetry can be inferred
through thermal transport measurements.
In the absence of broken time-reversal symmetry, we would anticipate
a vanishing thermopower $S_{xx}$ while observing a large Nernst signal $S_{xy}$
when the chemical potential is at the Dirac point \cite{Zhu2010,Proskurin2013}.
Conversely, if there is a chemical potential shift,
and the shift exceeds the Zeeman energy,
we expect to observe a large thermopower and a vanishing Nernst signal.
In the time-reversal symmetry broken state considered in this study,
there is asymmetry in the shifts of the two Dirac point energies
concerning the Fermi energy, along with a difference in their absolute values.
In this case, 
a pronounced Nernst signal and thermopower are anticipated,
with the Nernst signal exceeding the thermopower.

We compute $S_{xx}$ and $S_{xy}$ as follows:
In the presence of the electric field ${\bm E}$ and
the temperature gradient $\nabla T$,
the electric current is given by
\be
{\bm j}=\sigma {\bm E} + \alpha (-\nabla T),
\ee
in the linear response regime.
Here, $\sigma$ and $\alpha$ are the electrical and thermoelectric conductivity tensors,
respectively.
We first compute the zero-temperature conductivity, $\sigma_{ij}(\varepsilon_F)$,
where $\varepsilon_F$ is the Fermi energy, using the Kubo formula.
Subsequently, we obtain $\sigma$ and $\alpha$ at temperature $T$
and with chemical potential $\mu$ using the following equations \cite{Jonson1984,Zhu2010}:
\[
  {\sigma _{ij}}\left( {T,\mu } \right) = \int {d\varepsilon } {\sigma _{ij}}\left( \varepsilon  \right)\left( { - \frac{{\partial f\left( \varepsilon  \right)}}{{\partial \varepsilon }}} \right),
  \]
  \[{\alpha _{ij}}\left( {T,\mu } \right) =  - \frac{1}{{eT}}\int {d\varepsilon } {\sigma _{ij}}\left( \varepsilon  \right)\left( {\varepsilon  - \mu } \right)\left( { - \frac{{\partial f\left( \varepsilon  \right)}}{{\partial \varepsilon }}} \right).
  \]
  Here,
  $f\left( \varepsilon  \right) = 1/\left[ {\exp \left( {\beta \left( {\varepsilon  - \mu } \right)} \right) + 1} \right]$
  is the Fermi distribution function
  and $\beta=1/(k_{\rm B} T)$ with $k_{\rm B}$ being the Boltzmann constant.
    For the calculation of the zero-temperature conductivity, $\sigma_{ij}(\varepsilon_F)$,
    we account for the broadening of Landau levels due to impurity scattering, as detailed in
    Eqs.~(\ref{eq:sxx0}) and (\ref{eq:sxy0}).

  We assume the Landau levels of massless Dirac fermions of the continuum model,
  with parameters such as the Fermi velocity $v_F$ and the energy of the Dirac point $E_D$.
  For the two Dirac points present, we represent $E_D$ for each Dirac point as
  $E_D^{(1)}$ and $E_D^{(2)}$.
  Although the Dirac cones in \alphaI are tilted \cite{Tajima2018},
  the effect of the tilt is merely to renormalize
  the Fermi velocity \cite{Morinari2009,Goerbig2008}.
  To account for the broadening, we assume the Lorentz function
  with energy-independent damping, $\Gamma$, for simplicity.
    The formulas for $\sigma_{ij}(\varepsilon_F)$,
    specific to a single valley of the Dirac fermion state, are as follows:
    \bea
        {\sigma _{xx}}\left( {{\varepsilon _F}} \right)
        &=& \frac{{\hbar {e^2}v_F^2}}{{4\ell _B^2}}\sum\limits_{n = 0}^\infty
        \sum\limits_{\sigma  =  \pm }
        \left[
          \frac{{\Gamma /\pi }}
               {{{{\left( {{\varepsilon _F} - {\varepsilon _{n + 1}} - {\varepsilon _\sigma }} \right)}^2}
                   + {\Gamma ^2}}}
               \right. \nonumber \\
               & & \hspace{-5em} \left. 
          + \frac{{\Gamma /\pi }}{{{{\left( {{\varepsilon _F} + {\varepsilon _{n + 1}} - {\varepsilon _\sigma }} \right)}^2} + {\Gamma ^2}}}
          \right]
        \nonumber \\
        & &  \hspace{-6em} \times \left[ {\frac{{\Gamma /\pi }}{{{{\left( {{\varepsilon _F} - {\varepsilon _n} - {\varepsilon _\sigma }} \right)}^2} + {\Gamma ^2}}} + \frac{{\Gamma /\pi }}{{{{\left( {{\varepsilon _F} + {\varepsilon _n} - {\varepsilon _\sigma }} \right)}^2} + {\Gamma ^2}}}} \right],
        \label{eq:sxx0}
        \eea
        \bea
            {\sigma _{xy}}\left( {{\varepsilon _F}} \right)
            &=& \frac{{{e^2}}}{h}\sum\limits_{n = 0}^\infty
            {\sum\limits_{\sigma  =  \pm }
              {\left( {n + \frac{1}{2}} \right)} }
            \left[
              \theta_{\Gamma} \left( {{\varepsilon _F} - {\varepsilon _{n + 1}}
                - {\varepsilon _\sigma }} \right)
              \right. \nonumber \\
              & &  \hspace{-1em} \left.
                + \theta_{\Gamma} \left( {{\varepsilon _F} + {\varepsilon _{n + 1}}
                  - {\varepsilon _\sigma }} \right)
              \right.
              \nonumber \\
              & &  \hspace{-1em} 
              \left. { - \theta_{\Gamma} \left( {{\varepsilon _F} - {\varepsilon _n}
                  - {\varepsilon _\sigma }} \right)
                - \theta_{\Gamma} \left( {{\varepsilon _F} + {\varepsilon _n}
                  - {\varepsilon _\sigma }} \right)}
              \right].
            \label{eq:sxy0}            
            \eea
            Here, $\ell _B = \sqrt{\hbar/(eB)}$ is the magnetic length,
            and the function
            ${\theta _\Gamma }\left( x \right)$
            is defined by the equation below:
            \be
              {\theta _\Gamma }\left( x \right)
              = \frac{1}{2} + \frac{1}{\pi }{\tan ^{ - 1}}\left( {\frac{x}{\Gamma }} \right).
              \ee
            Including $E_D$
            in these formulas, the total values of 
            $\sigma_{ij}(\varepsilon_F)$
            for the two Dirac cones are represented as
            $\sigma_{ij}(\varepsilon_F+E_D^{(1)})
            +\sigma_{ij}(\varepsilon_F+E_D^{(2)})$
  In this simplified calculation, the peak
  in the longitudinal conductivity $\sigma_{xx}$ at $\varepsilon_F=0$ is underestimated,
  but there is no qualitative change
  when compared to the more elaborate calculation \cite{Zhu2010}.
  $S_{xx}$ and $S_{xy}$ are calculated using
  \be
  S_{ij}=\left( \sigma^{-1} \alpha \right)_{ij}.
  \ee

\begin{figure}[htbp]
  \includegraphics[width=\linewidth]{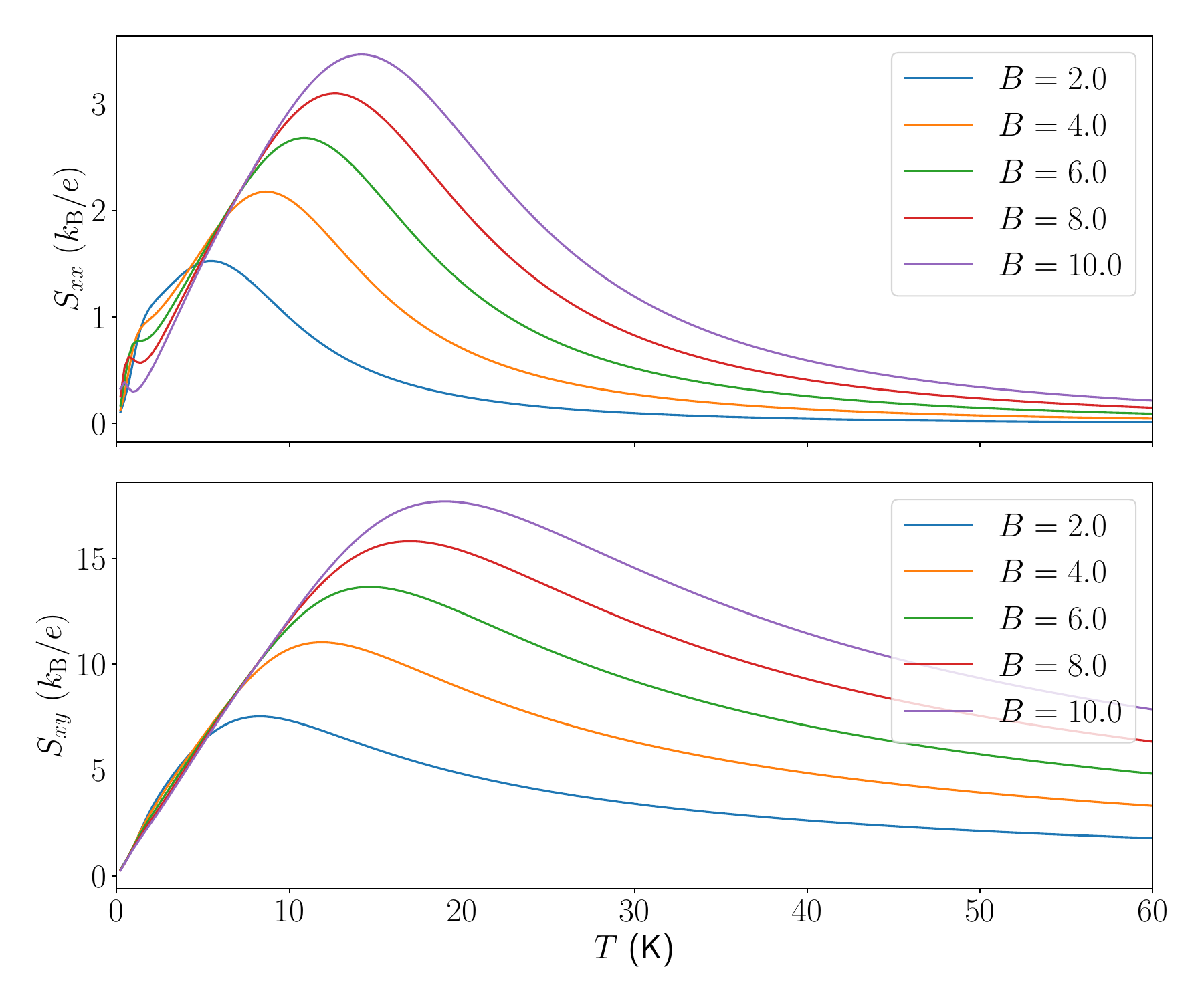}
  \caption{
    \label{fig:Sxx_Sxy}
    (Color online)
    The thermopower $S_{xx}$ (upper panel)
    and the Nernst signal $S_{xy}$ (lower panel) as
    functions of temperature for different magnetic fields,
    with the magnetic field measured in tesla.
    Both $S_{xx}$ and $S_{xy}$ are expressed in units of
    $k_{\rm B}/e=86.17~\mu$V~K$^{-1}$.
    We take $E_D^{(1)} = 0.8~$meV, $E_D^{(2)} = -0.7~$meV,
    and $\Gamma=0.1$~meV.
  }
\end{figure}

Figure~\ref{fig:Sxx_Sxy} presents the temperature dependence of $S_{xx}$
and $S_{xy}$ for different magnetic fields.
The crucial point to note is that both $S_{xx}$ and $S_{xy}$
exhibit large values, with $S_{xy}$ being several times larger than $S_{xx}$.
Here, we set the Fermi velocity $v_F = 5 \times 10^4 \, {\rm m} \, {\rm s}^{-1}$
for both Dirac cones.
The values for $E_D^{(1)}$ and $E_D^{(2)}$ are $E_D^{(1)} = 0.8$ meV
and $E_D^{(2)} = -0.7$ meV, respectively.
The Landau level broadening parameter is taken as $\Gamma = 0.1$ meV.

In the experiment\cite{Konoike2013}, the temperature dependence 
of the Nernst signal $S_{xy}$ displays a peak 
around $T \simeq 10$ K under magnetic field.
This peak emerges when the magnetic field exceeds 0.5 T, 
with its maximum observed around 10 T. 
This maximum value is notably large, surpassing 3 mV/K at 13 T and 11 K. 
Concurrently, the temperature dependence of the thermopower $S_{xx}$ 
also exhibits a peak around $T \simeq 10$ K 
under magnetic field. 
The peak value of $S_{xx}$, measured at 13 T and 11 K, is less than
one-sixth that of $S_{xy}$. 
These characteristics are consistent with our theoretical results.
In our computations, the primary parameters are $E_D^{(1)}$ 
and $E_D^{(2)}$.
The values specified above closely replicate the peak positions
and magnitudes of $S_{xy}$ and $S_{xx}$.
When we take either $E_D^{(1)} = -E_D^{(2)} \neq 0$ or $E_D^{(1)} = E_D^{(2)} = 0$,
$S_{xx}$ vanishes.
Meanwhile, if we take $E_D^{(1)} = E_D^{(2)} \neq 0$,
$S_{xy}$ is much smaller than $S_{xx}$.
To be consistent with the experimental observations, 
we require opposite signs and disparate absolute values for 
$E_D^{(1)}$ and $E_D^{(2)}$. 
This distinction is a direct result of our symmetry-broken phase.
A limitation of our calculation lies in the assumption
of a constant $\Gamma$.
Indeed, to account for the diminishing peak values
in $S_{xy}$ and $S_{xx}$ at high magnetic fields\cite{Konoike2013},
we must consider its variability.
This aspect will be addressed in subsequent research.

\section{Conclusion}
In conclusion, our study reveals that the ground state of \alphaI under pressure
breaks both time-reversal symmetry and particle-hole symmetry
while maintaining translational symmetry.
This symmetry breaking arises from the creation of fluxes
that deviate from 0 or $\pm \pi$ within a unit cell.
To experimentally confirm this symmetry breaking state,
we have calculated the thermopower and the Nernst signal,
and our results are in good agreement
with the experimental observations \cite{Konoike2013}.
This agreement provides substantial support for our theoretical prediction
regarding the symmetry broken state.

\begin{acknowledgments}
  The author thanks N. Tajima and T. Konoike for valuable discussions
  and to M. Ogata for providing insightful comments.
  This work was supported by JSPS KAKENHI Grant Number 22K03533.
\end{acknowledgments}

\bibliography{../../../../refs/lib}
\end{document}